\documentstyle[12pt, epsfig]{article}
\begin{document}


\begin{center}
{\Large \bf Detecting supersymmetric dark matter in M31 \\ with CELESTE ?}
\end{center}
\vspace{1.0 cm}

\begin{center}
{\large E.~Nuss\footnote{GAM, UMR5139-UM2/IN2P3-CNRS, Place Eug\`ene Bataillon, 34095 Montpellier, France}
G.~Moultaka\footnote{LPMT,UMR5825-UM2/CNRS, 
 Place Eug\`ene Bataillon, 34095 Montpellier Cedex 5, France}
A. Falvard${}^1$, E. Giraud${}^1$, 
A. Jacholkowska${}^1$, K. Jedamzik${}^2$, J. Lavalle${}^1$,
F.~Piron${}^1$, 
M. Sapinski${}^1$ 
P. Salati${}^3$, R.~Taillet\footnote{LAPTH, Annecy--le--Vieux, 74941, France}}
\end{center} 

%
%
\index{Falvard, A.}
\index{Giraud, E.}
\index{Jacholkowska, A.}
\index{Lavalle, J.}
\index{Nuss, E.}
\index{Piron, F.}
\index{Sapinski, M.}
\index{Salati, P.}
\index{Taillet, R.}
\index{Jedamzik, K.}
\index{Moultaka, G.}


%
\vspace{1.0 cm}

\begin{abstract}
It is widely believed that dark matter exists within galaxies and clusters
of galaxies. Under the assumption that this dark matter is composed of the 
lightest, stable supersymmetric particle, assumed to be the neutralino,
 the feasibility of its indirect detection via observations of a diffuse 
gamma-ray signal due to neutralino annihilation within M31 is examined. 
\end{abstract}
\vspace{3.0 cm}

\begin{center}
{\sl To appear in the proceedings of the French Astrophysical Society, 2002}
\end{center}
\newpage
%
\newcommand{\lsim}{\raisebox{-0.13cm}{~\shortstack{$<$ \\[-0.07cm] $\sim$}}~}
\newcommand{\gsim}{\raisebox{-0.13cm}{~\shortstack{$>$ \\[-0.07cm] $\sim$}}~}
\section{Introduction}
The existence of cosmic dark matter (DM) is required by a multitude of 
observations such as excessive peculiar velocities of 
galaxies within clusters of galaxies or gravitational arcs. 
Furthermore, both Big Bang nucleosynthesis (which predicts a baryonic relic 
density
$\Omega_b \ll \Omega_{tot} \simeq 1 $) and plausible scenario for large-scale structure formation, 
strongly suggest a substantial non-baryonic/cold DM component
in the Universe. It so happens that supersymmetric extensions of the standard 
model of particle physics provide a natural candidate for such a DM in the 
form of a stable uncharged Majorana
fermion (Neutralino). Hereafter, we briefly report on the potential of the high-energy $\gamma$--rays 
($E_\gamma \gsim 50$ GeV) ground based detector CELESTE (de Naurois {\sl et al.} '02),
to detect neutralinos indirectly through their annihilation in the
halo of M31. (For the detailed study see Falvard {\sl et al.} '02)

\section{Neutralino halo around M31 -- Neutralino annihilation}
 \noindent
The late-type Sb spiral galaxy M31 lying at a distance of 700 kpc
has a visible part consisting mostly of a bulge and a disk.
We have reconsidered the two mass components fit to the rotation
curve of M31, performed by Braun ('91). Taking the following mass-to-light
ratios $\Upsilon_{bulge} = 6.5 \pm 0.4~\Upsilon_{B,\odot}$ 
and $\Upsilon_{disk} = 6.4 \pm 0.4 ~\Upsilon_{B,\odot}$
(where $\Upsilon_{B,\odot}$ is the mass-to-light ratio for the Sun)
Braun concluded that no dark halo is necessary to account for the
velocity field. However, this conclusion relies on a large
value of $\Upsilon_{disk}$ which disagrees with estimates based on the 
blue color of the disk and on synthetic spectra of young stellar
populations which it contains (Guiderdoni '87). 
The mass-to-light ratio $\Upsilon_{disk}$ of a purely stellar component
should actually not exceed 
$\sim$ $3.8~\Upsilon_{B,\odot}$. 
In addition, a disk as massive as that proposed by Braun
should generally be unstable.
Therefore, we have assumed the presence of an additional
mass component in terms of a spherical halo whose mass density
profile is generically given by
\begin{equation}
\rho_{\chi}(r) \; = \; \rho_0 \,
\left( \frac{r_0}{r} \right)^\gamma \,
\left\{ {\displaystyle
\frac{r_0^\alpha + a^\alpha}{r^\alpha + a^\alpha}}
\right\}^\epsilon \;\; 
\label{neutralino_rho}
\end{equation}

\noindent 
where $\gamma, \alpha$ and $\epsilon$ define the various profiles. 
While a fit to the observations constrains weakly the ratio 
$\Upsilon_{disk}/\Upsilon_{bulge}$, it leads to more stringent
constraints on the structure of the neutralino halo, actually favouring
a NFW profile $\gamma = 1, \alpha = 1,\epsilon =2$
(Navarro, Frenk, \& White '96). Typical results are illustrated in
Fig. 1.

\begin{figure}[h]
   \centering
\vspace{-0.5cm}
   \epsfig{file=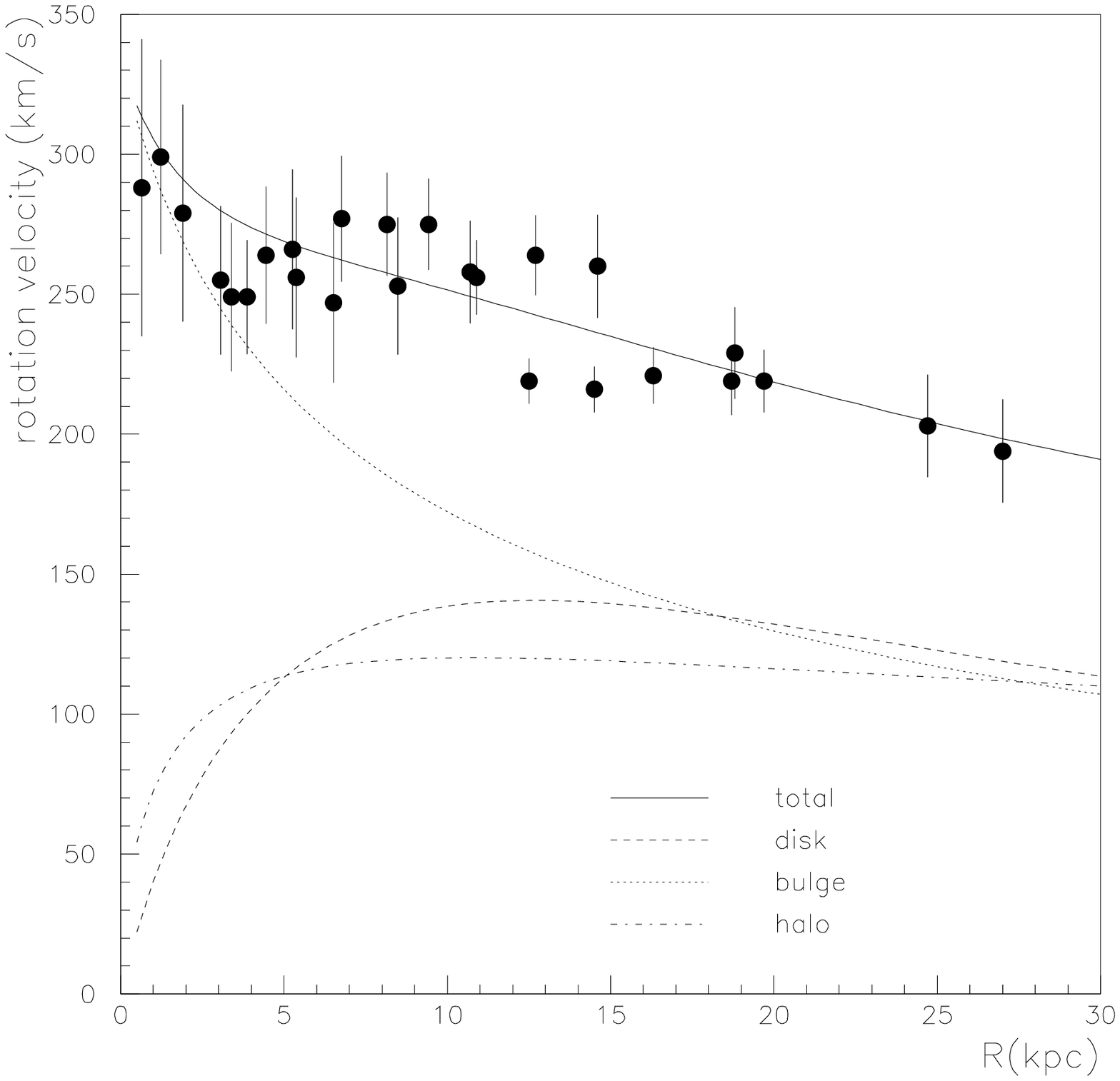, width=6cm}
\vspace{-0.5cm}
    \epsfig{file=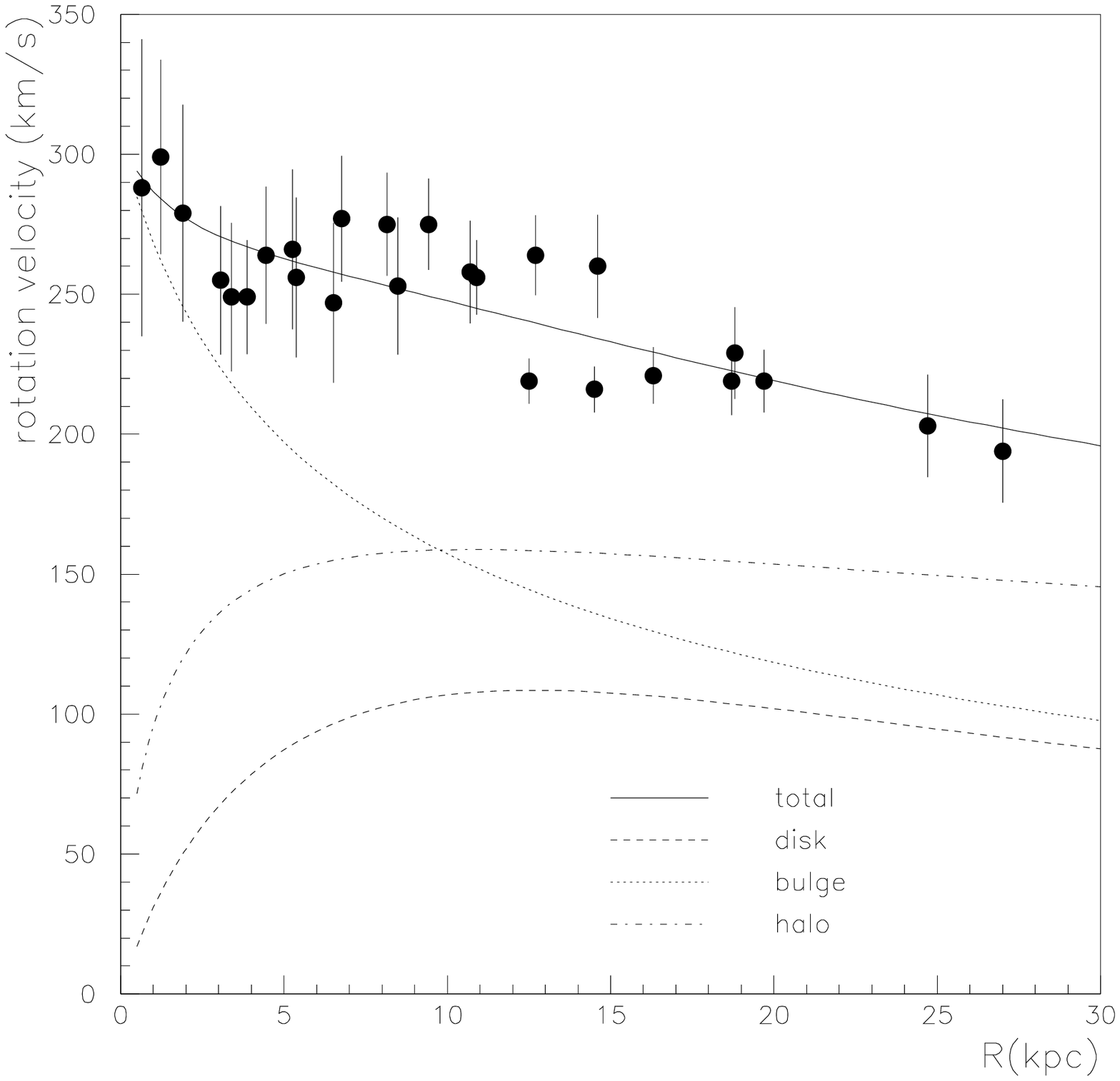, width=6cm}
      \caption{ A $\gamma = 1$ neutralino halo is added to the bulge and to the
disk of M31. {\sl Left:} an intermediate case 
with  $\Upsilon_{bulge} = 4.2 ~\Upsilon_{B,\odot}$
and $\Upsilon_{disk} = 4.2~\Upsilon_{B,\odot}$. 
{\sl Right:}  a maximal
halo with $\Upsilon_{bulge} = 3.5~\Upsilon_{B,\odot}$, 
$\Upsilon_{disk} = 2.5~\Upsilon_{B,\odot}$. The global solid  rotation
curve is in good agreement with the data of Braun.}
       \label{figure_RT3}
   \end{figure}
\noindent
If neutralinos are indeed a substantial component of the DM
around M31, one expects them to annihilate into standard model particles, 
eventually producing energetic photons.
The corresponding photon flux at Earth -- per unit of time and surface -- 
may be expressed as
\begin{equation}
I_\gamma \; = \; \frac{dn_\gamma}{dt\,dS}\; = \; \frac{1}{4 \pi} \,
{\displaystyle
\frac{\langle \sigma v \rangle \, N_{\gamma}}{m_{\chi}^{2}} } \,
{\displaystyle \int_{\rm fov} \int}_{\rm los} \rho_{\chi}^{2} \, ds \,d\Omega\;\; 
\equiv \; \frac{1}{4 \pi} \,
{\displaystyle
\frac{\langle \sigma v \rangle \, N_{\gamma}}{m_{\chi}^{2}} } \,
\Sigma\,
\label{gr_flux_1}
\end{equation}
where $m_{\chi}$ is the neutralino mass and
$\langle \sigma v \rangle \, N_{\gamma}$ denotes the thermally averaged 
annihilation rate yielding $N_{\gamma}$ gamma-rays in the final state.
Eq.(\ref{gr_flux_1}) encapsulates all the particle physics features
 in $\langle \sigma v \rangle \, N_{\gamma}/m_{\chi}$,   
while the astrophysical modelling is contained in the line of sight
integral $\Sigma$. 
In Table 1 of Fig.3  we illustrate some typical fluxes taking a nominal
neutralino mass, $m_{\chi} = 500$ GeV, and annihilation cross section,
$\langle \sigma v \rangle \, N_{\gamma} \, = \, 
{10^{-25} cm^3 s^{-1}}$.  
\section{Supersymmetric model predictions and resulting fluxes} 
In the present section we consider more specific particle physics model predictions
of the $\gamma$--ray fluxes. We will focus mainly on the minimal supergravity scenario 
(mSUGRA),
(Barbieri {\sl et al.} '82, Chamseddine {\sl et al.} '82,
Hall {\sl et al.} '83). Making the usual simplifying assumption of
 common universal values at some grand unified theory (GUT)
 energy scale $M_{GUT} \sim 2 \times 10^{16}$GeV, {\sl i.e.} $ m_{scalars}(M_{GUT}) \equiv m_0, 
M_{gauginos}(M_{GUT}) \equiv m_{1/2}, A_{trilinear}(M_{GUT}) \equiv A_0$, 
and requiring various physical consistencies (e.g. electroweak symmetry breaking at low scale, 
electrically neutral and stable DM particles, present experimental limits
from particle colliders, etc...) one can make specific predictions for the $\gamma$ flux
in terms of these model parameters. Assuming the neutralinos should account for a large
fraction of the DM at cosmological scales as well, one also requires the
corresponding relic density to be in the right hallmark, $0.025 \lsim \Omega_{\chi} h^{2} \lsim 0.3$.
In Fig. 2 we present a scan over the basic parameters of mSUGRA for the correlation between
the integrated $\gamma$--flux and the neutralino mass and relic density.
All results have been obtained using an interfaced version of the two public codes DarkSUSY and
SUSPECT. Convoluting the annihilation spectra with
the CELESTE acceptance, we show in Fig.3 the expected rate of $\gamma$/min from a maximal
M31 smooth halo (see Fig.1). This puts the signal from such halos clearly beyond a reasonable CELESTE
reach which is roughly one tenth of the CRAB nebula (4.9 $\gamma$/min).
Nonetheless, a significant departure from a smooth halo is not unrealistic due to possible
clumpiness and/or black hole accretion effects. This can lead to a global enhancement factor 
which may reach up to two orders of magnitude, as illustrated in Table 2.,
 making the survey
of M31 with CELESTE worth being undertaken.

\section{Conclusion}
We conclude that under favourable conditions such as 
rapid accretion of the neutralinos on the central black hole in M31
and/or excessive halo clumpiness, a neutralino annihilation $\gamma$-ray
 signal may be seen by the ongoing observations of M31 with CELESTE.

\begin{figure}[h]
   \centering
\vspace{-0.5cm}
   \epsfig{file=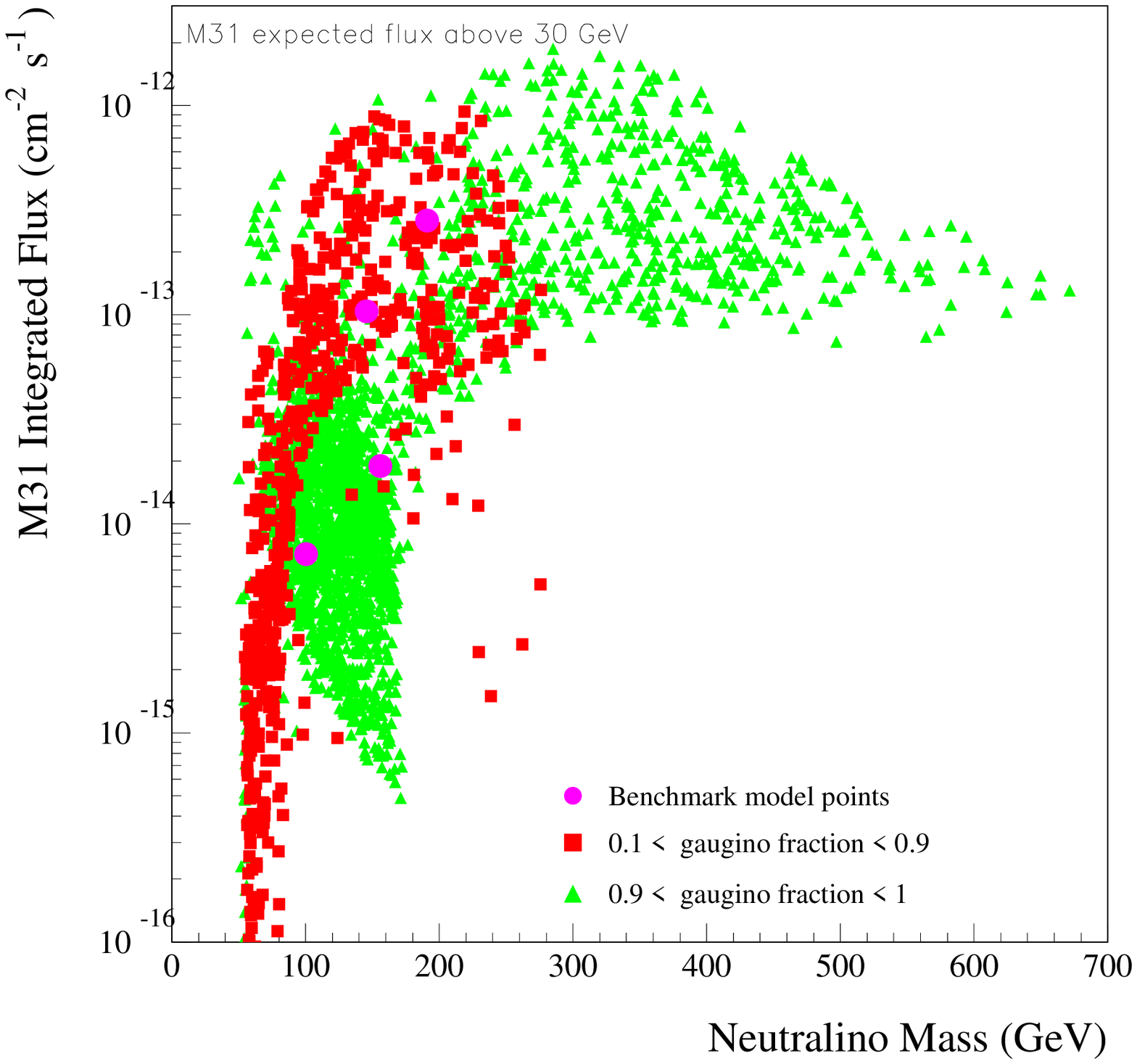, width=6cm}
\vspace{-0.5cm}
   \epsfig{file=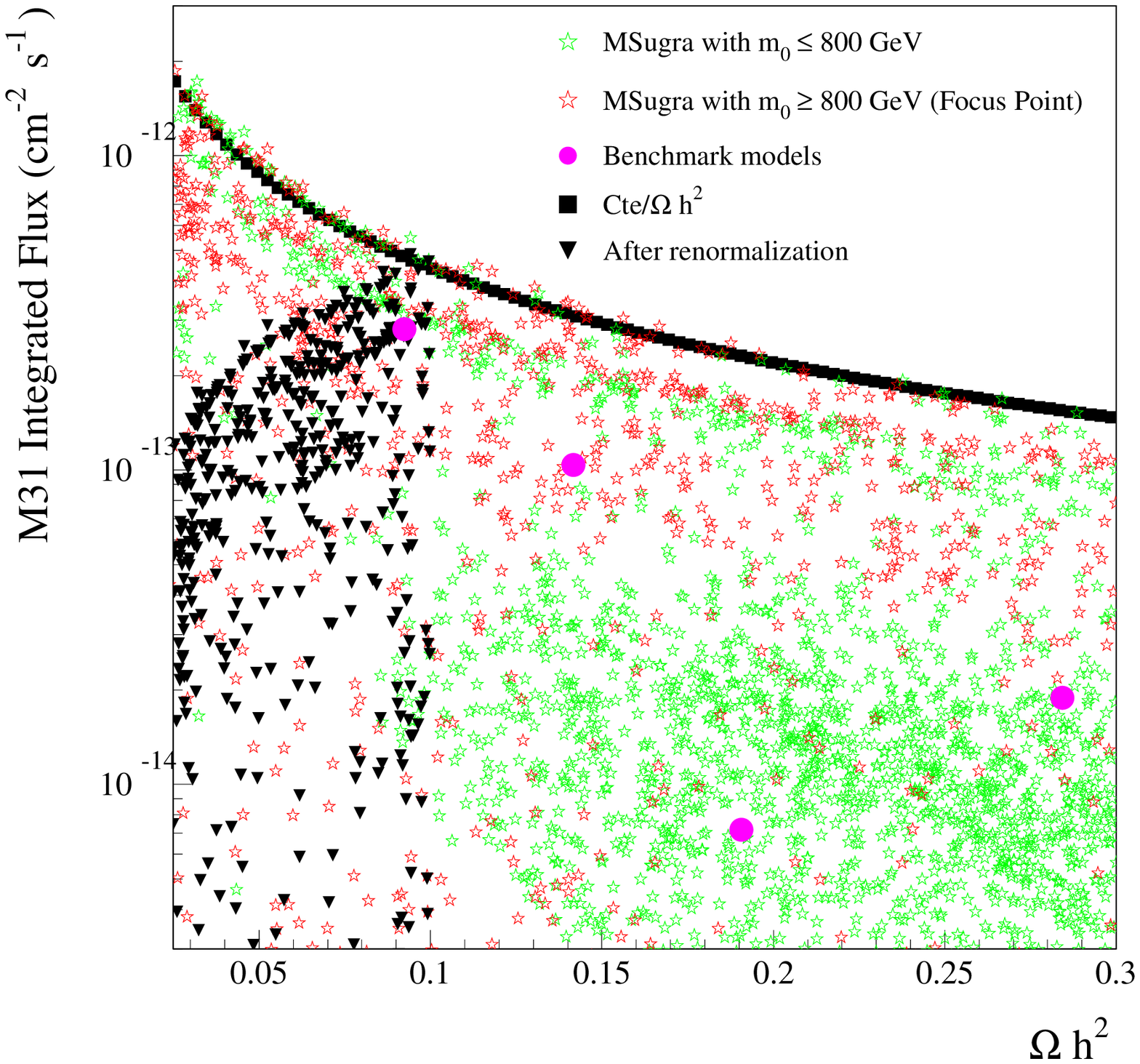, width=6cm}
   \caption{{\sl Left}: the integrated $\gamma$ flux from M31 as a function of 
$m_{\chi}$ for $E_{\gamma} > 30$ GeV. Each point corresponds to a model in our "wild scan". 
Three different ranges of gaugino fraction are considered.
{\sl Right}: the integrated $\gamma$ flux from M31 as a function of
$\Omega_{\chi}$$h^{2}$. A renormalization procedure for 
$\Omega_{\chi}$$h^{2} < 0.1$ has been applied on flux values (see Falvard A. {\sl et al.} '02). }
   \centering
     \vspace{-.5cm}
     \epsfig{file=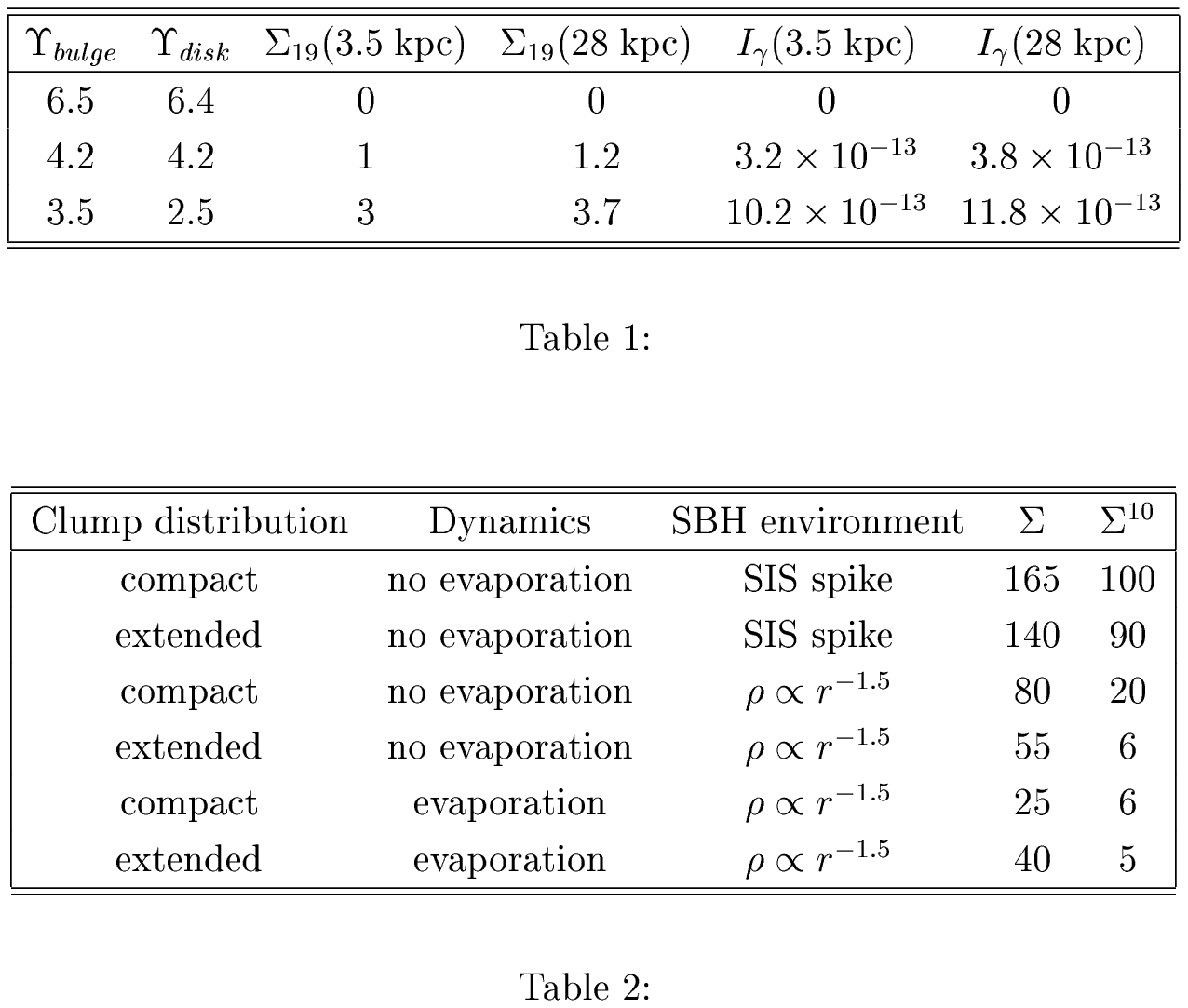, width=6cm}
         \epsfig{file=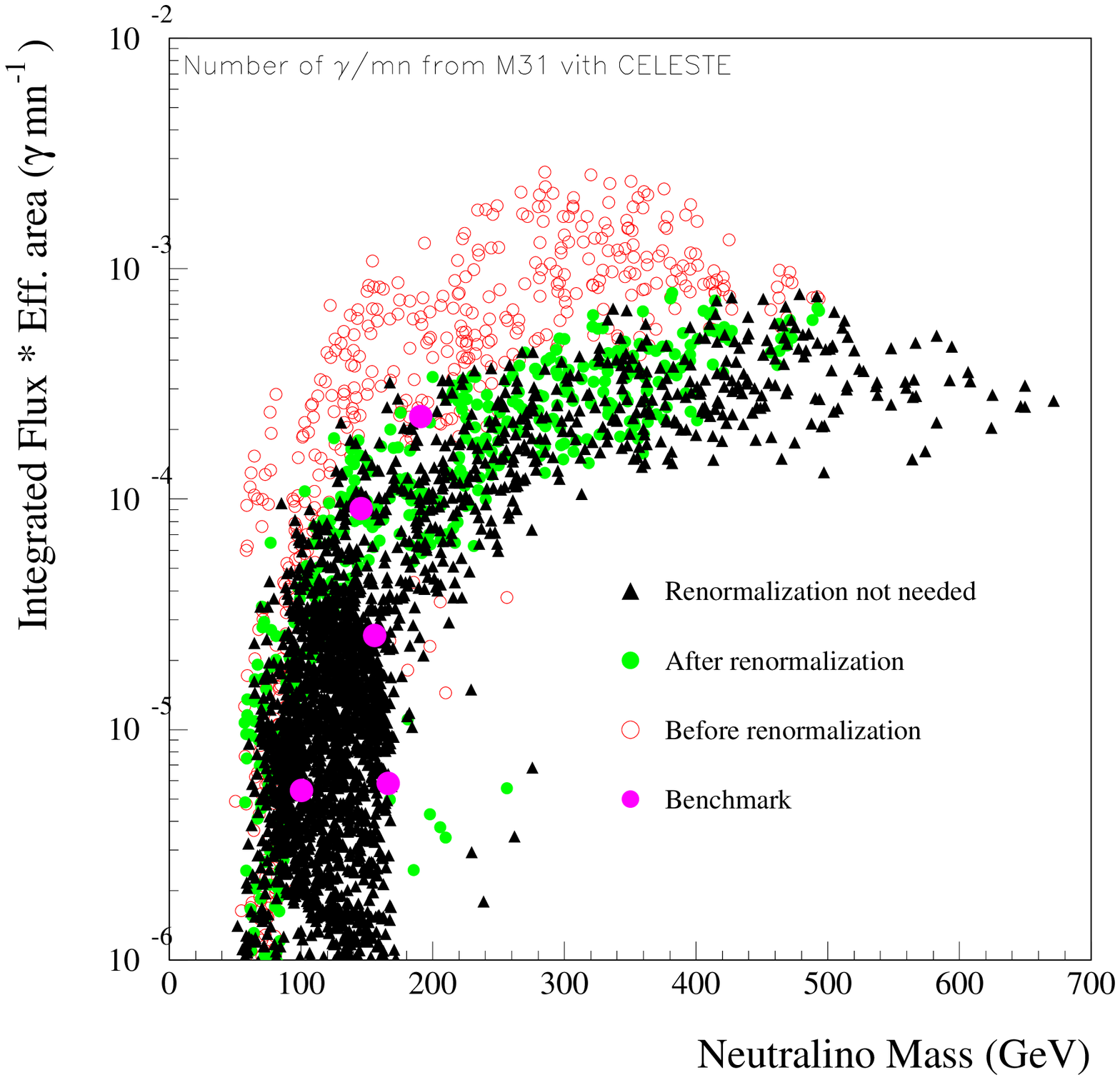, width=6cm}
    \caption{Table 1: Three different models for M31 are featured.
The first corresponds to the case where no halo is needed. 
$\Sigma_{19}(R)$ is in units of $10^{19}$$GeV^{2} cm^{-5}$ and the flux $I_{\gamma}$(R)
is for a circular region encompassing the inner 3.5 kpc 
(corresponding to the CELESTE f.o.v. of $10$ mrad)
 and 28 kpc, respectively.
Table 2: Impact of astrophysical parameters such as clumpiness of
the M31 halo and supermassive black hole (SBH) in its centre, on flux predictions, 
smooth and clump contributions added. ($\Sigma^{10}$ corresponds to the CELESTE f.o.v.).
{\sl Right}: Number of $\gamma / min$ as expected with CELESTE from M31. }
    \end{figure}


\end{document}